# The architect Kha's protractor

Amelia Carolina Sparavigna
Dipartimento di Fisica,
Politecnico di Torino, Torino, Italy

Kha was an architect at Deir El-Medina, Egypt, supervisor of some projects completed during the reigns of three kings of the 18$^{th}$ Dynasty (approximately 1440-1350 BC). Buried with his wife Merit, the items of their tomb are exposed at the Egyptian Museum, Torino. After a description of some devices of the ancient Egypt masonry (cubits, cords, plumb, levels and squares), that Kha used during his activity, I discuss an object among those found in his tomb, which, in my opinion, could be used a protractor.

In 1906, Arthur Weigall and Ernesto Schiaparelli discovered the Theban Tomb n.8 (TT8), on behalf of the Italian archaeological mission [1,2]. Considered as one of the greatest archaeological discoveries concerning the ancient Egypt, this tomb of the New Kingdom survived intact till its discovery. The pyramid-chapel of Kha and his wife Merit had already been well-known for many years, as shown by some paintings of this chapel reproduced by Karl Lepsius (1810–1884), but the tomb was located far from the chapel. Egyptologists also knew that Kha was an important supervisor at Deir El-Medina, responsible for some projects completed during the reigns of kings, Amenhotep II, Thutmose IV and Amenhotep III [1,3].

The coffins and objects from TT8 tomb are now at the Museo Egizio of Turin. The tomb held the funerary equipment of Kha the architect and of his wife Merit. The items found in the tomb show Kha and Merit as quite wealthy people (very delicate are the Merit's objects for her afterlife). Moreover, included in one of Kha's coffins was one of the earliest examples of the Egyptian Book of the Dead. Kha had, among his items, two cubits. One was gilded and inscribed, probably a royal gift to the architect. The other cubit was of wood, and could be folded by hinges. We can suppose that this was used by the architect in his activity [4]. As told in Ref.4, some cubits found in the tombs are of wood, the material actually used for the cubit rods by architects, some were votives or specifically made of hard materials for the funerary equipment, or presents, such as that gilded with gold, found in the Kha's Tomb.

Let us see the devices at disposal of Kha, during his activity of architect. In the ancient Egypt masonry, several measuring devices were used: cubits, cords, plumb, levels and squares. The cubits resemble our rulers, about 52.5 centimetres long. The front faces were used for marking and numbering for measurements. The rear faces were often inscribed with the name of the owner. The cubit was subdivided in 7 palms of 4 digits [4,5]. It seems that stonemasons used another measuring unit, the nby-rod, about 67 to 68 centimeters long, subdivided into seven spaces [4]. For measuring great distances, Egyptians used ropes, as we use the measuring tapes, probably with a length of 100 cubits (52.5 meters). Some pictures show the architects working with ropes having knots, to indicate the subdivisions [4].

To have the vertical and horizontal directions, the Egyptian used plumbs and levels. Plumbs consisted simply of a plumb bob suspended from pegs or sticks. The plumb bobs vary in material, shape, and size, frequently in the shape of amulets. According to [4], Petrie collected a great variety, with very old examples from the Third Dynasty. An elaborate type of plumb was among the burial objects in the tomb of the architect Senedjem (approximately, 1280-1220 BC) at Deir el-Medina [4]. Moreover, a plumb used in a square level device was found in this tomb too. These level devices had probably the shape of the letter A. The string of the plumb was fixed at the top of the two legs, that is, at the top of the letter. The tip of the plumb had to touch the center of the cross board.

Egyptian builders and masons made use of simpler wooden devices, the squares, in order to have right angles. In principle, these instruments are the same that we use today. These levelling instruments of ancient Egypt continued to be used in Roman and medieval masonry as well. They were superseded by the water level, known to the Romans, but used for special purposes [4].

Of course, we can suppose that Kha had a balance too. The oldest known balances are the equal-armed ones, found in Egypt and are represented on Egyptian reliefs and drawings, often depicted in the Books of Dead, in the scene of the "Weighing of the Heart" in front of the afterlife tribunal. According to Ref.6, the existence of a balance standard in the Fifth Dynasty is testified. To see the use of plumb in the balance scales, to check its arms horizontality, the reader can see images and discussions in Ref.7.

One of the objects from the Kha's Tomb, Egyptian Museum of Torino, is supposed to be the case of a balance scale (see Fig.1), or the scale itself. This is what we read from the label. In a previous preparation of the items of Kha's Tomb, it was possible to see the front and back of the object (see Fig.2). They are the same, with the same complex decoration. In my opinion, this decoration had a functional use too, may be as a protractor, to determine directions. Let us then observed the decoration in detail (see Fig.3); the decoration has a 16-fold symmetry, looking as a compass rose with 16 leaves. But, outside this rose there is a polygonal line with 18 corners, pointing outwards. They correspond to the same number of corners, but inwards. That is, we have a line with 36 corners.

In the inner decoration we can see the fraction 1/16 corresponding to one leaf. The Egyptian knew and used the fractions as the sum of distinct unit fractions. That is, a fraction was written as a sum of fractions, each fraction having a numerator equal to 1 and a denominator equal to a positive integer. Every positive rational number can be represented by an Egyptian fraction [8]. In this ancient system of calculus, the Eye Of Horus defined the Old Kingdom number one, such as, 1 = 1/2 + 1/4 + 1/8 + 1/16 + 1/32 + 1/64, rounded-off at the six-term. A 1/64 is needed to have the exact value 1. The separated parts of the Eye of Horus were used to write describe the fractions [9]. May be, Kha used the balance case for calculations, or simply as a protractor when he was using a plumb, a level or a scale, to find vertical or equilibrium positions, but also to measure the deviation from vertical or horizontal directions.

Is there anything we can tell about number 36, the number of corners of the external decoration of the architect's case? Let us remember that the Egyptians has the Decans, 36 groups of stars which rise in succession from the horizon due to the earth rotation. The rising of each decan marked the beginning of hours of the night. In the ancient Egypt, the decans were used as a sidereal clock beginning by at least the 9th or 10th Dynasty (ca 2100 BC) [10].

Probably, the object found in the Kha's tomb had simply a geometrical decoration, such as that on another object of the tomb, which seems to be a "rose of direction" (Fig.4). But, the use of 1/16 fraction, the coincidence of the number of corners with that of decans, and the fact that the decoration was engraved on the instrument of an architect, suggest me that this object had been used as a protractor instrument with two scales, one based on Egyptian fractions, the other based on decans.

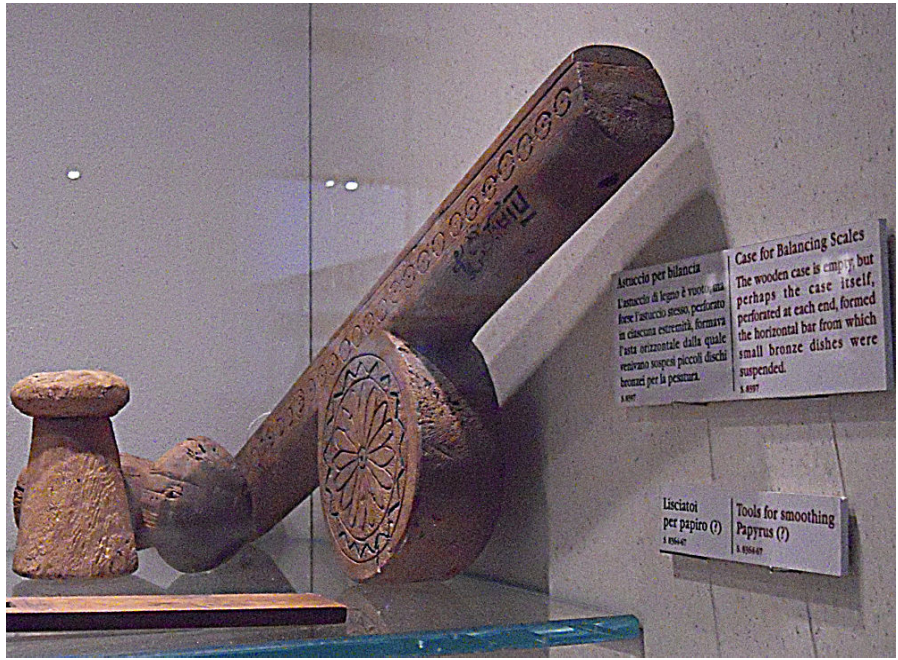

Fig.1 The label is telling that the object is the case of a balance scale (Egyptian Museum, Torino)

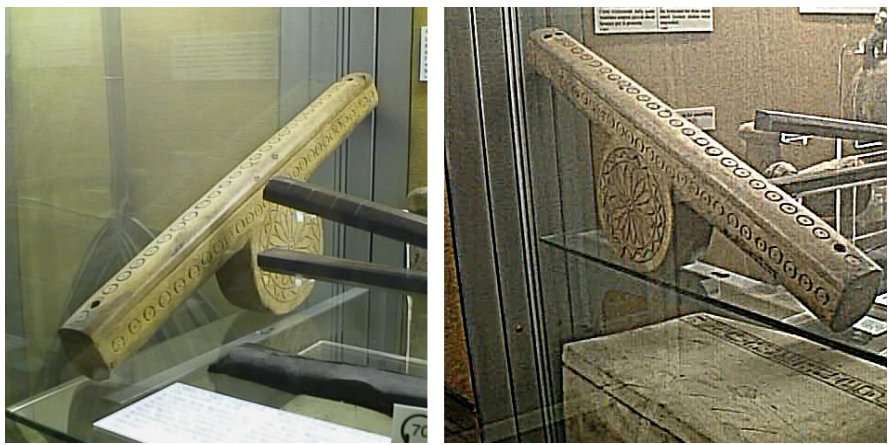

Fig.2 In a previous preparation of the items found in the Kha's Tomb at the Egyptian Museum, it was possible to see the front and back of the object. They are the same.

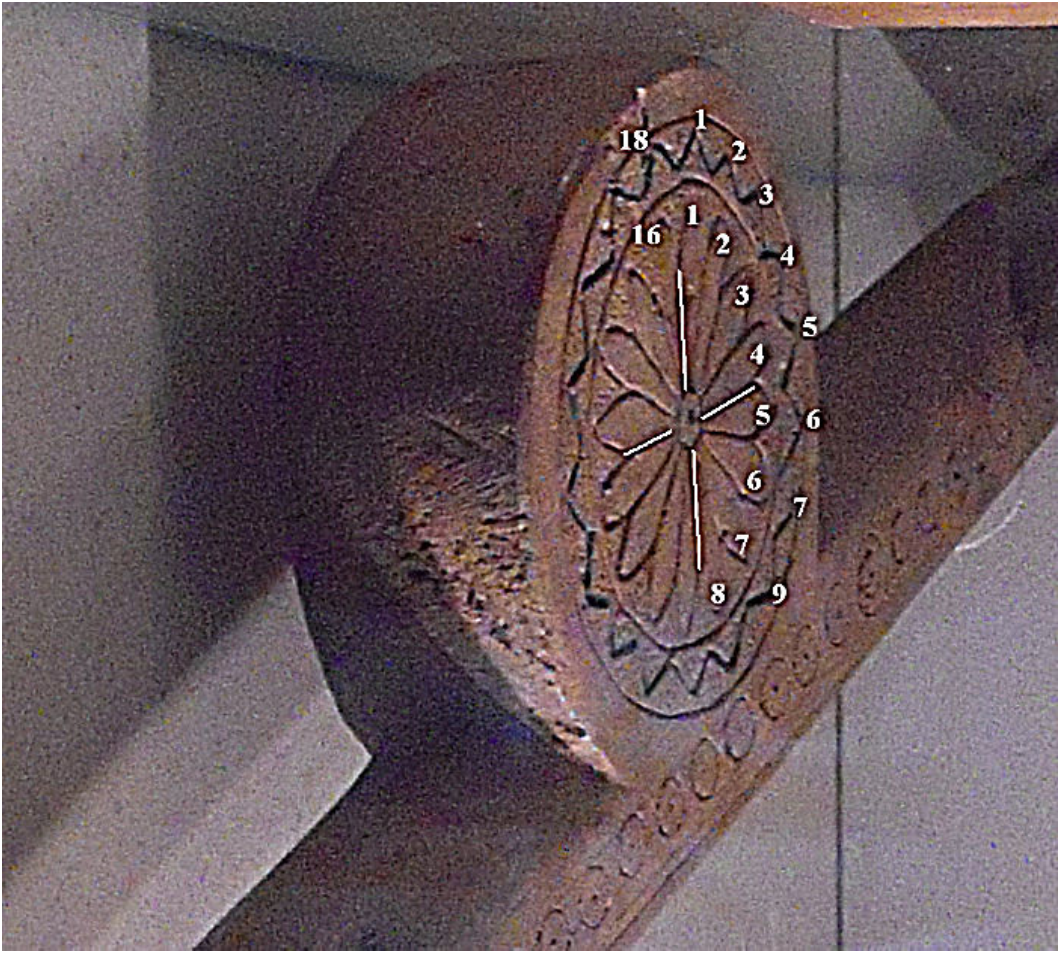

Fig.3 The decoration of the object of Kha is quite complex: it has 16-fold symmetry, looking as a compass rose with 16 leaves. Outside this rose, there is a polygonal line 18 corners, pointing outwards. They correspond to the same number of corners (pointing inwards). That is we have a line with 36 corners.

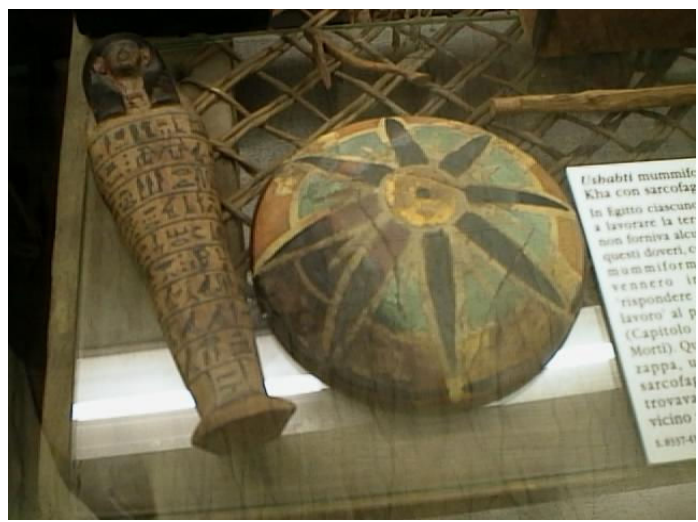

Fig.4 A "rose of direction", from the Kha's Tomb.